\documentclass{emulateapj}
\usepackage{textcomp}
\usepackage{mathtext}
\usepackage{amssymb}
\usepackage{rotating}
\usepackage{diagbox}

\newcommand{\fermi}{\textit{Fermi}}
\newcommand{\gr}{$\gamma$-ray}
\newcommand{\lsi}{LS I  $+$61\textdegree303}

\begin{document}

\title{Superorbital modulation at GeV energies in the $\gamma$-ray binary \lsi}

\author{Yi Xing, Zhongxiang Wang}
\affil{Shanghai Astronomical Observatory, Chinese Academy of Sciences,
80 Nandan Road, Shanghai 200030, China}
\author{Jumpei Takata}
\affil{Physics Department, Huazhong University of Science and Technology, Wuhan 430074, China}

\begin{abstract}
We report the results from our analysis of 8 years of the data 
for the $\gamma$-ray binary \lsi,
obtained with the Large Area Telescope onboard the {\it Fermi Gamma-Ray Space
Telescope}. We find a significant dip around the binary's periastron in
the superorbital light curves, and by fitting the light curves with a
sinusoidal function, clear phase shifts are obtained. The superorbital
modulation seen in the binary has been long known and different scenarios
have been proposed. Based on our results, we suggest that 
the circumstellar disk around the Be companion of this binary may
have a non-axisymmetric structure, which 
rotates at the superorbital period of 1667 days. As a result, 
the density of the ambient
material around the compact star of the binary changes along the binary orbit
over the superorbital period, causing the phase shifts in the modulation, 
and around periastron, the compact star probably enters the Be disk or
switches the mode of its emission due to the intereaction with the disk, 
causing the appearance of the dip. We discuss the implications of
this possible scenario to the observed superorbital properties at
multiple frequencies.

\end{abstract}


\section{Introduction}
Galactic X-ray binaries consist of a neutron star or a black hole as the
primary and a companion star. They are luminous X-ray sources in the sky
powered by accretion with mass tranferred from the companion to the primary. 
In addition to the often-seen periodic signals related to their binary 
orbits, long-term, so-called
superorbital periodic signals are also seen in a few of them. The origin
of the superorbital flux variations is not clearly determined, but is thought
to be due to the procession or warping of the accretion disk around a compact
star (see \citealt{kc12} for a brief review and references therein). 
Other proposed possible origins include X-ray state changes, 
the existence of precessing jets,
or the existence of a tertiary star around a binary.

Having a Be star as the companion, \lsi\ is one of a handful high-mass X-ray 
binaries (HMXBs) that are \gr\ loud 
(see, e.g., \citealt{dub13}).  Such \gr\ binaries are of great interests and 
have been under extensive studies, as complicated physical proecesses
occur in them and produce remarkable phenomena at multiple wavelengths.
The orbit of \lsi\ is elliptical with a period $P_o$ of 
26.496 days \citep{gre02}
and an eccetricity of $e\simeq 0.54$ \citep{ara+09}. Orbital flux or
emission line variations have been
detected at all wavelengths, from radio (e.g., \citealt{per90,mar+16}), 
optical (e.g., \citealt{gru+07,zam+14}), X-ray 
(e.g., \citealt{par+97}), to \gr\ GeV and TeV 
(e.g., \citealt{abd+09,alb+09}). Very interestingly, a 
long-term 1667$\pm$8 days periodic signal \citep{gre02} is also seen 
at nearly all the wavelengths (radio: \citealt{gre02,mt16}; optical:
\citealt{zm00};
X-ray: \citealt{che+12,sah+16}; GeV: \citealt{ack+13}; TeV: \citealt{ahn+16}).
Note that \citet{mt16} recently have determined
a value of 1628$\pm48$ days for the long-term period.

How the superorbital varations are produced in \lsi\ is under debate. 
It has been thought that the variations are related to the cyclical changes
in the mass loss of the Be companion or the density/size of 
the circumstellar disk around the companion \citep{zam+99}.  
The interaction of the compact star (which is
most likely a neutron star in this case; e.g., \citealt{ptr12,tor+12})
with the disk is modulated, giving rise to
the variations. Recently, arguing that the superorbital modulation is stable
in the long-term radio data, \citet{mt14} have proposed a jet model, in which
the jets from an accreting black hole precess at a period $P_2$ slightly 
longer than $P_o$, $P_2\simeq 26.9$ days and the beat frequency corresponds to
the superorbital signal.

In order to fully investigate the superorbital variations at $\gamma$-ray
GeV energies and help understand the origin, we conducted detailed 
analysis of the data obtained with
the Large Area Telescope (LAT) onboard the {\it Fermi Gamma-Ray Space
Telescope (Fermi)} using the latest Pass 8 database. Previously \citet{ack+13}
have found sinusoidal variations, prominently seen during apastron orbital 
phases,
at the superorbital period of 1667 days from the LAT data. \citet{sah+16}
recently have also obtained similar variations from their analysis of nearly
7-years LAT Pass 8 data. In addition to these results, our analysis indicates
a significant dip around periastron
and a clear phase-shift trend in the orbital-phase-resolved superorbital
modulations. 
To explain the features, we suggest that the circumstellar disk around
the companion would be eccentric and precess at the superorbital period.
Here we report the results.

\section{Data Analysis and Results} 
\label{sec:ana}

\subsection{\textit{Fermi} LAT Data}

LAT, onboard \fermi, is a $\gamma$-ray imaging instrument 
that scans the whole sky every three hours and can continuously 
observe thousands of GeV sources in the sky \citep{atw+09}. 
In this analysis, we selected 0.1--300 GeV LAT events 
from the \textit{Fermi} Pass 8 database inside 
a $\mathrm{20^{o}\times20^{o}}$ region, which is centered at the position 
of \lsi\ given in the \fermi\ LAT third source catalog \citep{ace+15}. 
The time period of the LAT data is 8 years
from 2008-08-04 15:43:36 (UTC) to 2016-08-18 00:48:16 (UTC). Following 
the recommendations of the LAT 
team\footnote{\footnotesize http://fermi.gsfc.nasa.gov/ssc/data/analysis/scitools/}, 
we included events with zenith angles less than 90 degrees 
(preventing the Earth's limb contamination) and 
excluded the events with quality flags of `bad'.

\subsection{Source's \gr\ Properties and Orbital Variability}

We first repeated the likelihood analysis and spectral analysis to
the whole selected data,
and confirmed the previously reported results obtained for this source
\citep{abd+09,sah+16}. In addition, we also repeated orbital variability
analysis by folding the source's photons at the orbital period and obtaining
the spectra during the periastron and apastron phase ranges.
\begin{figure}
\centering
\epsscale{1.2}
\plotone{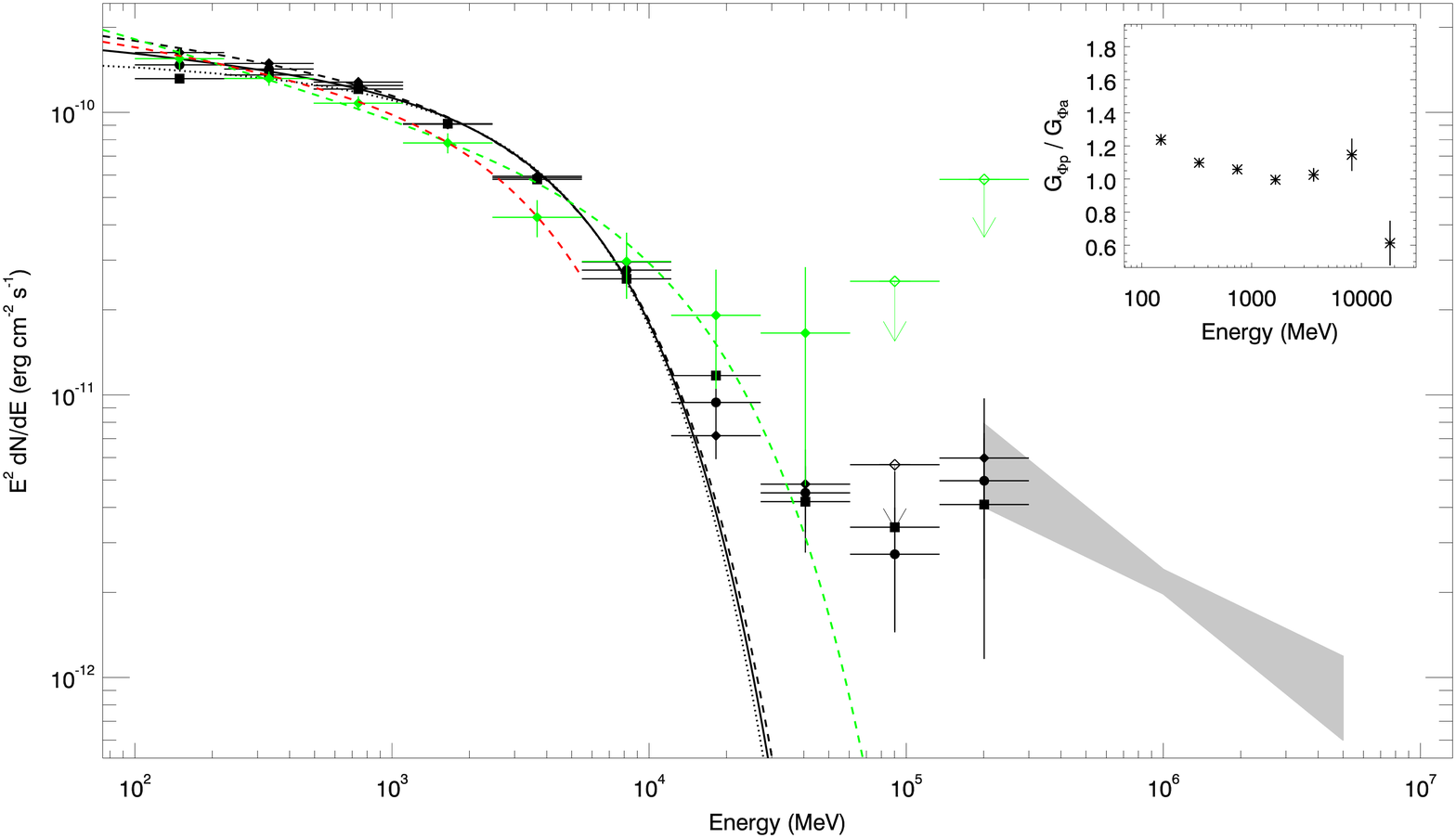}
\caption{\fermi\ \gr\ spectra of \lsi\ during 
$\Phi_{t}$ (circles), $\Phi_{p}$ (diamonds) and $\Phi_{a}$ (squares). 
The black solid, dashed, and dotted lines are the 0.1--300 GeV 
exponentially cutoff power-law fits during $\Phi_{t}$, 
$\Phi_{p}$, and $\Phi_{a}$, respectively. 
The grey area marks the power-law spectrum of \lsi\ obtained with 
VERITAS \citep{ali+13}, which was detected during the time intervals 
approximately within $\Phi_{a}$. The ratios of the energy fluxes 
between $\Phi_{p}$ and $\Phi_{a}$ are plotted in the inner panel.
The green data points are the fluxes obtained during the superorbital
dip, and the green and red dashed lines are the exponentially cutoff power-law 
fits to the dip data at 0.1--300 and 0.1--5.5 GeV, respectively (see
the text in Section~\ref{sec:dip}).
}
\label{fig:spectra}
\end{figure}

\subsubsection{Likelihood analysis}
\label{sec:la}

We included all sources within 20 degrees centered at the position 
of \lsi\ in the \textit{Fermi} LAT 4-year catalog \citep{ace+15} to make 
the source model. The spectral forms of these sources are provided in 
the catalog. The spectral parameters of the sources within 5 degrees 
from \lsi\ were set as free parameters, and the other parameters were 
fixed at their catalog values. \lsi\ was included in the source model 
as a point source with emission modelled with an exponentially cutoff 
power law, $dN/dE = N_{0}E^{-\Gamma}\exp[-(E/E_{c})]$, 
where $\Gamma$ and E$_{c}$ are the photon index and cutoff energy, respectively.
In addition, 
the background Galactic and extragalactic diffuse emission was included,
with the spectral model gll\_iem\_v06.fits and the 
file iso\_P8R2\_SOURCE\_V6\_v06.txt, respectively, used
in the source model.  The normalizations of 
the diffuse components were set as free parameters.

Using the LAT science tools software package {\tt v10r0p5}, we performed 
standard binned likelihood analysis to the LAT data 
in the 0.1--300 GeV band. 
The Instrument Response Functions (IRFs) of P8R2\_SOURCE\_V6 were 
used.
The \gr\ emission of \lsi\ during the total data set ($\Phi_{t}$) was 
detected to have a Test Statistic (TS) value of 136681, and 
$\Gamma= 2.086\pm$0.009 and $E_{c}= 5.5\pm$0.2 GeV were obtained.
For comparison, we also repeated the likelihood analysis  
with emission from \lsi\ modelled with a simple power law 
$dN/dE = N_{0}E^{-\Gamma}$, and obtained $\Gamma= 2.314\pm$0.004 
with a TS value of 140997. 
We estimated the significance of the spectral cutoff from 
$\sqrt{-2\log(L_{pl}/L_{exp})}$, where
$L_{exp}$ and $L_{pl}$ are the maximum likelihood values for  
the power law model with and without the cutoff 
respectively \citep{abd+13}. 
The spectral cutoff was detected  
at a significance of 38$\sigma$. The parameters of the exponentially cutoff 
power-law fit are listed in Table~\ref{tab:likelihood}.

\subsubsection{Spectral analysis}
\label{sec:saf}

We extracted the $\gamma$-ray spectrum of \lsi\ by performing maximum 
likelihood analysis of the LAT data in 10 evenly 
divided energy bands in logarithm from 0.1--300 GeV. 
In the extraction, the spectral normalizations of the sources within 5 
degrees from \lsi\ were set as free parameters, while all the other parameters 
of the sources were fixed at the values obtained from the above maximum 
likelihood analysis. We kept only spectral data points when TS greater 
than 9 ($>$3$\sigma$ significance) and derived 95\% flux upper limits 
otherwise. The obtained spectrum is shown in Figure~\ref{fig:spectra}, with
the flux values of the spectral data points provided 
in Table~\ref{tab:spectra}. 

\subsubsection{Orbital variability analysis}
\label{sec:ova}

\begin{figure}
\centering
\epsscale{1.0}
\plotone{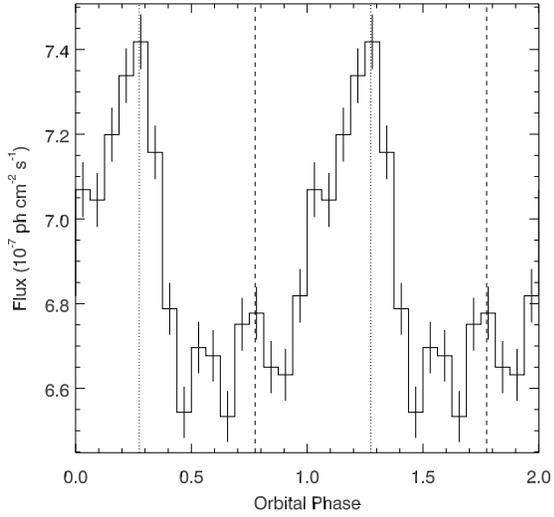}
\caption{
Orbital light curve of \lsi, where the dotted and dashed lines mark 
the phases of the periastron and apastron, respectively. For clarity,
two cycles are displayed.
}
\label{fig:orbital}
\end{figure}

We performed timing analysis to the 0.1--300 GeV LAT data of \lsi\ to 
study the orbital variations of the source. We selected \gr\ photons 
within an aperture radius of 2 degrees from \lsi, as most of the sources 
in the source model are located $>$2 degrees away from \lsi, and weighted 
them by their probabilities of originating from the binary 
(calculated with \textit{gtsrcprob} in the science tools software package) 
using the fitted source model obtained in Section~\ref{sec:la}. 
Orbital phases $\phi_{o}$ for the photons were assigned using the ephemeris 
given in \citet{ara+09}. We folded the weighted photons into 16 orbital 
phase bins, and plotted them over the exposures (calculated 
with \textit{gtexposure}) in each of the bins in Figure~\ref{fig:orbital}. 
The folded light curve has modulation peak approximately at the periastron 
($\phi_{o}$ of 0.275), consistent with that in the previous studies 
(\citealt{abd+09,had+12}).

We further defined the orbital phase ranges of 0.0--0.5 around periastron 
and 0.5--1.0 around apastron as $\Phi_{p}$ and $\Phi_{a}$, respectively, 
and performed likelihood analysis and spectral analysis to the LAT data 
during these two phase ranges. The likelihood results are given in 
Table~\ref{tab:likelihood}. The spectra are plotted in 
Figure~\ref{fig:spectra}, with the spectral flux values listed in 
Table~\ref{tab:spectra}. The $\Gamma$ value during $\Phi_{p}$ is $>$6$\sigma$ 
higher than that during $\Phi_{a}$ (see Table~\ref{tab:likelihood}), 
indicating a softer \gr\ emission during the former than the latter. 
The $E_{c}$ value during the two phase ranges are consistent within 
uncertainties. In order to show the detailed differences between the emission
from the two phase ranges, we compared their ratios of the energy fluxes 
(the inner panel of Figure~\ref{fig:spectra}, where
only the ratios in the energy ranges of $<$27.2 GeV are shown as the flux
uncertainties in the energy ranges of 27.2--60.5 and 134.7--300~GeV are too
large to be reliable).
As can be seen, the \gr\ emission during the former is higher than 
that during the latter in the low energy ranges, and the ratio decreases 
to $\simeq 1$ in the middle energies, and in the high energy range 
of 12.2--27.2 GeV the \gr\ emission during the former is lower than 
that during the latter. 
This comparison helps indicate the spectral changes between the two phase
ranges found from the likelihood analysis.

\subsection{Superorbital Variability Analysis}
\label{sec:sva}

\subsubsection{Orbitally-resolved superorbital variability}
\label{sec:osv}

From previous work \citep{ack+13,sah+16}, it has been shown that
the shape of superorbital modulation depends on orbital phase.
We therefore performed the same superorbital analysis.
According to the ephemeris given in \citet{gre02}, we selected LAT photons in 10 superorbital phase bins and performed likelihood analysis to the data in each of the bins. The photons in 10 orbital phase
ranges (0.0--0.1, ..., 0.9--1.0) were separatedly considered.
We note that the superorbital period
in \citet{gre02}
is consistent with that found by \citet{mt16} within the uncertainties.
In addition the period difference is only $\sim$0.025 in phase (which is small 
compared to the 10 superorbital phase bins). We tested by using
the latter period in analysis and found that the difference 
did not affect our results obtained in the following analysis.

\begin{figure*}
\centering
\epsscale{1.0}
\plotone{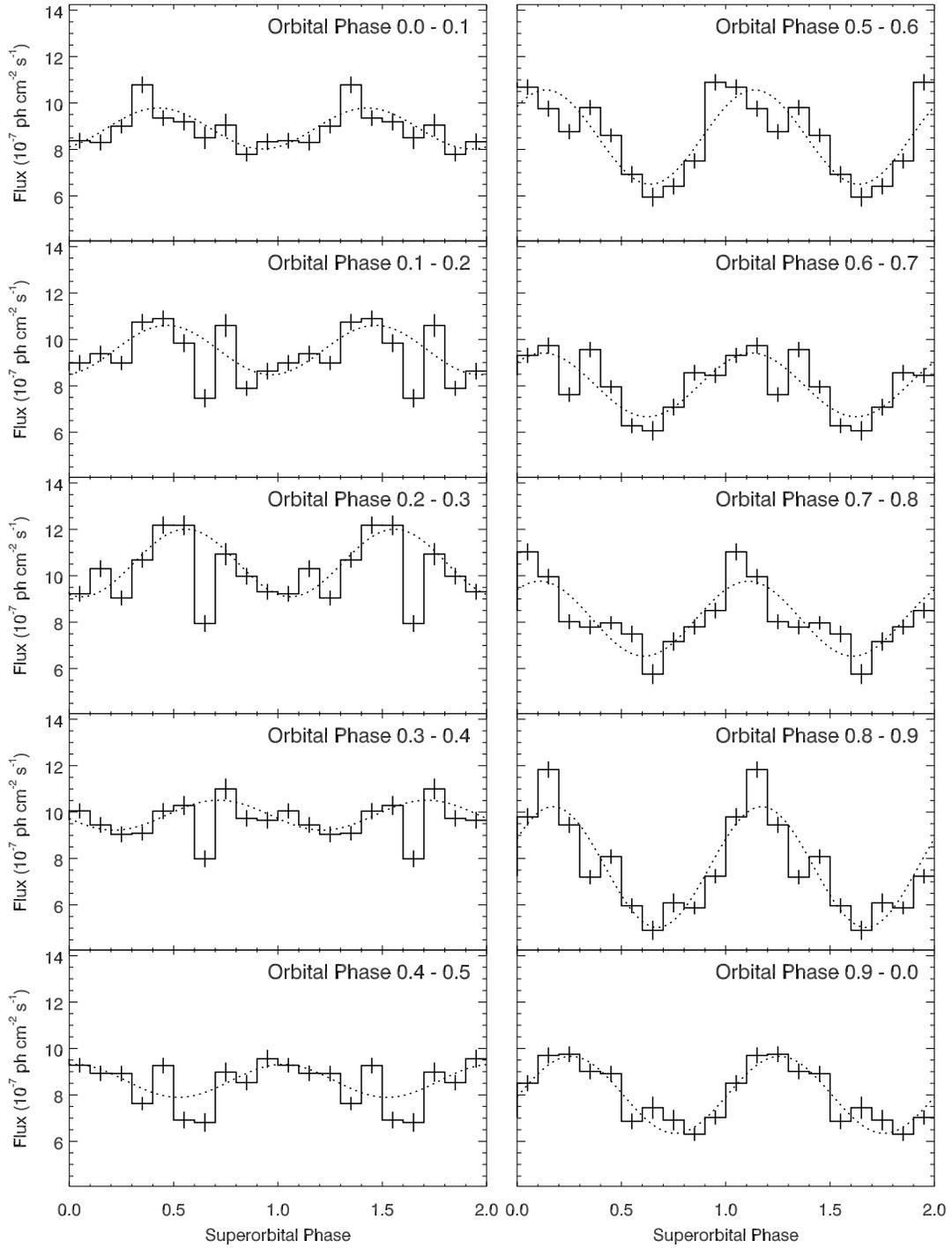}
\caption{0.1--300 GeV superorbital light curves in 10 orbital phase ranges, with the best-fit sinusoidal functions (Table~\ref{tab:fit_orbital_resolved_superorbital}) plotted as dotted lines. Two cycles are displayed for clarity.
}
\label{fig:orbital_resolved_superorbital}
\end{figure*}

The 10 light curves are shown
in Figure~\ref{fig:orbital_resolved_superorbital}. As can be seen,
the superorbital modulations were significant during the five orbital phase
ranges 
near apastron: the $\chi^{2}$-test values are 128--332 for 9 
degrees of freedom (dof; a value of 22 corresponds to a 99\% confidence 
level for variations). During the other five orbital ranges near periastron, 
the modulations are relatively weak (the $\chi^{2}$-test values are 42--120 for 9 
dof), but with a significant dip appearing at superorbital phase
$\phi_{so}$= 0.65 in the orbtial ranges of 0.1--0.4. 
Note that this dip occurs near the superorbital phase where the modulation
peaks are located at (see the light curves in the orbital phase ranges of
0.1--0.4).
Excluding the dip point, 
we checked the $\chi^{2}$-test values again for the three light curves, 
$\chi^{2}=$21--84 
for 8 dof (a value of 20 corresponds to the 99\% confidence level), indicating 
that the superorbital modulations were still significant.

Most of the light curves appear to have a sinusoidal-like shape, and thus 
following the previous work in \citet{ack+13} and \citet{sah+16}, we fit 
them with a sinusoidal function 
$A[1+$sin$(2\pi(\phi_{so}-\phi_{so,peak})+\pi/2)]/2+C$, where $\phi_{so,peak}$ 
is the superorbital peak phase, and $A$ and $C$ are the modulation amplitude 
and constant flux (in units of photon cm$^{-2}$ s$^{-1}$), respectively. 
For the three light curves near periastron that have the dip, the dip data point 
was not included in the fitting. The resulting reduced $\chi^2$ values were
in a range of $\sim$1.5--10.4, suggesting that either most of the light curves
are not exactly sinusoidal or the uncertainties on the data points were 
under-estimated.
In any case, we added an artificial value in quadrature with the uncertainties
to lower the reduced $\chi^2$ values to $\sim$1.
The obtained best fits are plotted as dotted lines in 
Figure~\ref{fig:orbital_resolved_superorbital}, and the fit results are given 
in Table~\ref{tab:fit_orbital_resolved_superorbital}. We also evaluated 
the significance of the dip by comparing the observed and best-fit model 
fluxes at $\phi_{so}=$ 0.65, which are also listed in 
Table~\ref{tab:fit_orbital_resolved_superorbital}. The dip was indeed
significant during the orbital phase ranges of 0.1--0.4 ($\simeq $3--5$\sigma$).
During the other seven orbital phase ranges, the dip was not significantly 
detected.

\begin{figure}
\centering
\epsscale{1.1}
\plotone{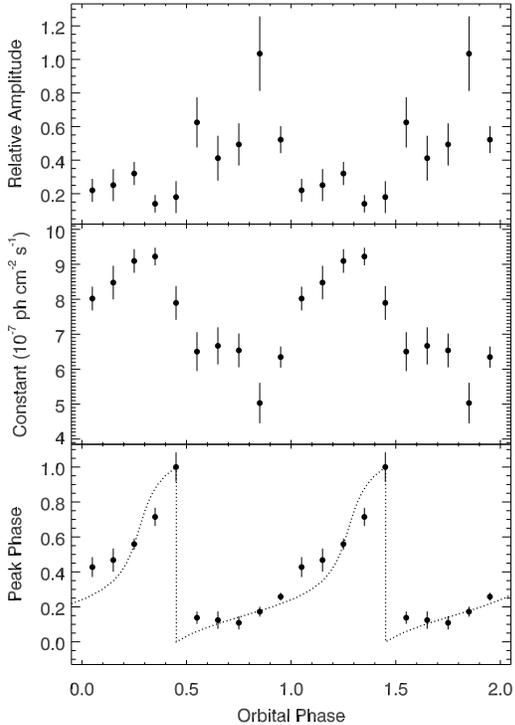}
\caption{Parameters of the best-fit sinusoidal functions for the superorbital light curves in each of the 10 orbital phase ranges. Two cycles are displayed for clarity. The dotted line in the bottom panel is the model curve obtained by
considering a precessing eccentric disk around the Be companion (see the text
in \S~\ref{sec:disc}).}
\label{fig:sofit}
\end{figure}

As pointed above, the superorbital modulation
amplitudes near apastron are obviously larger than those 
near periastron.  We show the best-fit parameters of the sinusoidal 
functions in 
Figure~\ref{fig:sofit}, and the results
confirm it quantatively: the relative amplitudes ($A/C$) are
approximately between 10\%--30\% during $\phi_{o}= 0.0$-0.5 and
40\%--100\% during $\phi_{o}=0.5$--1.0. On the other hand,
the constant fluxes near periastron are approximately
40\% higher than those near apastron. 
In addition, a striking feature is that the superorbital modulation peak 
shifts with orbital phase, from $\phi_{so}\simeq 0.1$ at 
$\phi_{o}=0.55$ to $\phi_{so}\simeq 1.0$ at $\phi_{o}=1.45$
(see the bottom panel of Figure~\ref{fig:sofit}).
This feature was noted by \citet{ack+13} and \citet{sah+16}, but our analysis
for the first time shows that there is a consistent pattern for 
the shifts over the whole superorbital/orbital phase.

\subsubsection{Properties of the dip}
\label{sec:dip}

\begin{figure}
\centering
\epsscale{1.2}
\plotone{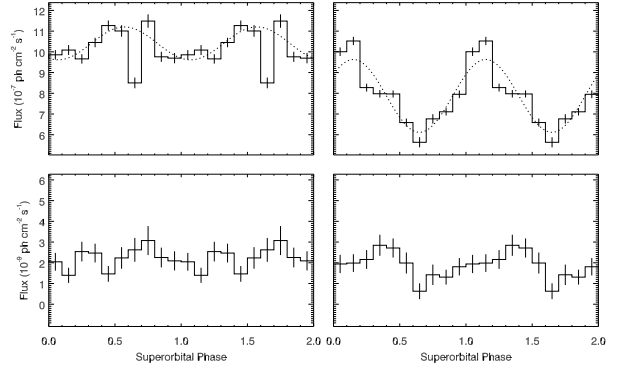}
\caption{Superorbital light curves in $\leq$5.5 ({\it upper}) 
and $>$5.5 GeV ({\it bottom}) bands during 
$\phi_{o}=0.135$--0.435 ({\it left} panels).
The corresponding light curves during $\phi_{o}=0.625$--0.925 
are shown in the {\it right} panels.
The best-fit sinusoidal functions to the light curves in
the upper panels are plotted as dotted lines. Two cycles are displayed
for clarity.
}
\label{fig:sodip}
\end{figure}

We investigated the dip's properties by first determining the orbital phase 
range in which the dip is the most significant.
We searched through different range values by repeating the above sinusoidal
fitting process, and found that in the 0.3 orbital phase range centered at 
$\phi_{o}$=0.285, the dip was detected at the highest significance of 
7$\sigma$ (after adding an artificial value to the uncertainties to
have a reduced $\chi^2\simeq 1$).  We also checked its
dependence on energy. Analyzing light curves in different energy ranges,
we found that the dip exists only significantly in the energy range 
of $<$5.5 GeV. 
The superorbital light curves in the energy ranges of below and above
5.5 GeV are shown in the left panels of 
Figure~\ref{fig:sodip}. 
In the $>$5.5 GeV energy range, the light curve was nearly a constant,
as the $\chi^2$ test value was only 12.0 (for 9 dof). As a comparison,
the 0.3 phase range light curves around the apastron were shown
in the right panels of the figure. For the $>$5.5 GeV one, 
modulation is still visible,
with a $\chi^2$ test value of 21.5 (for 9 dof; at a 98.9\% confidence level).

We also performed analysis to the data during the dip $\Phi_d$
determined above (i.e., $\phi_o=0.135$--0.435 and $\phi_{so} =0.6$--0.7),
in order to check if there are any physical differences between the dip
and the other phase ranges. 
The likelihood results are given in Table~\ref{tab:likelihood}, and the obtained
spectrum is shown in Figure~\ref{fig:spectra}, with the spectral flux values 
listed in Table~\ref{tab:spectra}. The photon index, $\Gamma=2.27\pm0.02$,
is significantly higher than
those obtained from the total data or periastron/apastron orbital phase
ranges and the cutoff energy, $E_c=16\pm4$~GeV, is also higher
but with the very large uncertainty. Examining the dip spectrum 
in Figure~\ref{fig:spectra}, the spectral data points may not be well described
by the exponentially cutoff power-law fit, as the fit has a higher tail 
(resulting the high cutoff energy of 16 GeV). 
Since the dip is only significant
in the $<$5.5 GeV data, we tested to limit the likelihood analysis to 
the 0.1--5.5 GeV data and found that the fit results, $\Gamma=2.15\pm0.03$ and
$E_c=4.2\pm0.5$~GeV (see also Table~\ref{tab:likelihood}) are then consistent.
Therefore there is no clear evidence for having a different spectrum during
the dip.

\section{Discussion}
\label{sec:disc}

At \fermi\ LAT energies, \gr\ emission from \lsi\ obviously have two 
components, one described by a power law with an exponential cutoff at
$\sim 5$ GeV and one slightly rising from $\sim 27$ GeV to 200~GeV.
The former and the latter can be connected to the emission at
MeV/KeV and TeV energies respectively (e.g., \citealt{znc10}). 
No matter whether the primary is a pulsar or a black hole with jets,
the former is likely due to synchrotron radiation and the latter due to
the inverse Compton (IC) scattering (e.g., \citealt{znc10,sah+16}).
This spectral feature is similar to that in the high-mass pulsar binary
PSR~B1259$-$63/LS~2883 (e.g., \citealt{xwt16}), the only \gr\ binary with
a known primary for a long time, although the synchrotron tail
of the latter source approximately ends at $\sim 1$ GeV.
(We note that in either pulsar or black hole accretion models such as 
those in \citealt{znc10} or \citealt{jtm16}, respectively,
GeV emission can come from the IC scattering processes, but then the TeV
component remains to be explained.)

Based on the above general scenario for emission from \lsi, our analysis 
has shown that in addition
to be orbitally modulated, the synchrotron component is also superorbitally 
modulated, particularly by showing a dip that is significantly seen 
at $<5.5$~GeV energy range around periastron (cf., Figure~\ref{fig:sodip}).
Although
the modulations at different orbital phases may not be well described by
a simple sinusoid used in our analysis, the general properties we have
obtained should not be affected. Near periastron, the average flux is high and
the modulation amplitude is low, and near apastron, we see the opposite
(Figure~\ref{fig:sofit}).
The sinusoidal-like modulation feature has been pointed out by 
the previous studies \citep{ack+13,sah+16},
but our analysis not only has confirmed its existence over the whole orbital 
phase ranges, but also been able to, by detecting a dip near 
the periastron phases, better determine the modulation properties.

Previously the superorbital modulation is often discussed as the result
of a cyclical change in the density/size of the circumstellar disk of 
the masssive companion (e.g., \citealt{ack+13,sah+16}), as such changes 
have been 
relatively well observed in Be star systems (e.g., \citealt{riv13}), including
Be/X-ray binaries (e.g., \citealt{neg+01}). For the Be disk in \lsi,
changes in the radius of the disk at the superorbital period
has been clearly seen in long-term optical spectroscopy \citep{zm00,zam+13}.
However in order to explain the clear trend of the phase shifts
we have obtained (cf., Figure~\ref{fig:sofit}), 
a non-axisymmetric structure is needed (axisymmetric changes cannot produce
the phase shifts, which are a function of orbital phase). 
This structure would rotate at the long
period of 1667 days, inducing the observed phase shifts. 
A well-studied case for comparison is the superhump modulation
observed in white dwarf binaries and low-mass X-ray binaries 
(e.g., \citealt{pat+05,has+01}), which are caused
by the precession of an elliptical or warped accretion disk around 
the compact star (see, e.g., \citealt{wk91,has+01}). 

\begin{figure}
\centering
\epsscale{1.2}
\plotone{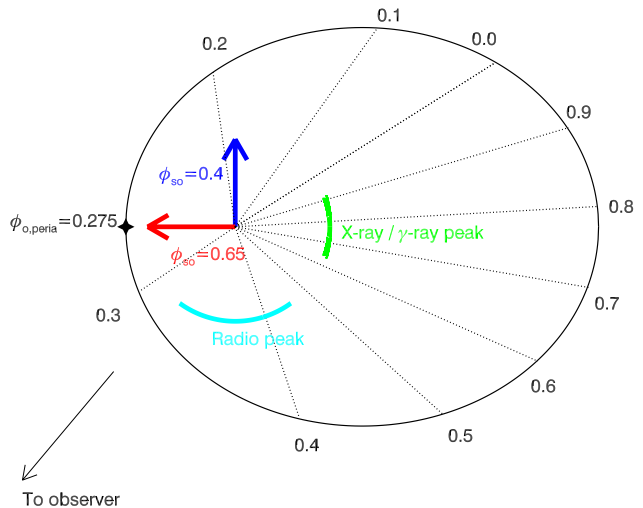}
\caption{Schemetic diagram for a non-axisymmetric Be disk precessing
at a period of 1667 days, where $\phi_{so}=0.65$ corresponds to when
the extended or denser part of the disk points to periastron (see details
in Section~\ref{sec:disc}). In this scenario, the precession 
of such a disk
would cause the superorbital modulation observed at radio, X-ray, 
and GeV \gr\ frequencies, whose peaks' superorbital phase ranges are 
indicataed for the discussion purpose.
}
\label{fig:orb}
\end{figure}

Here we suggest a similar scenario for \lsi. Since around periastron, 
the source's overall flux is high, particularly in
$<$1~GeV low-energy range (Figure~\ref{fig:spectra}), it is reasonable 
to assume that the synchrotron component is somehow related to 
the high density condition to the compact star. 
If the Be disk is not axisymmetric in its structure, for exmaple having 
an elliptical shape or a mode with density enhanced along one direction,
the preccession of such a disk would result in the density variations 
along the binary orbit. Assuming that the highest
density, which reflects the precession, leads to the \gr\ emission peak 
(see the schemetic diagram shown in 
Figure~\ref{fig:orb}), we obtain the dotted curve in
the bottom panel of Figure~\ref{fig:sofit}. The curve can be seen approximately
explains the phase shifts.  According to observations and related numerical
simulations, non-axisymmetric structures in Be disks in binary systems do 
exist (e.g., \citealt{riv13,pan+16}). Specifically in Be/X-ray binaries,
due to the tidal force of a neutron star, spiral density waves and an
eccentric mode (in eccentric binaries) in a Be disk can be 
developed \citep{oka+02}. In addition, in the numerical simulation 
designed for \lsi, long spiral density waves are seen due to the interaction
between the compact star and the disk, as when the compact star passes 
periastron, it tidally deforms the disk \citep{rom+07}.
Therefore it is possible that in addition to the long-term periodic 
changes in radius, the disk might have an elliptical shape or an eccentric 
denser part due
to the tidal force of the compact star and its precession would give rise
to the phase shift phenomenon. This possibility can be investigated by
numerical simulations designed to check if a non-axisymmetric pattern in
the Be disk of \lsi\ can be kept for a sufficiently long term (see
\citealt{oka+02} for their detection of a steady eccentric mode).

In this scenario, we have set $\phi_{so}=0.65$ at periastron
(Figure~\ref{fig:orb}), 
the reason for which is to explain the dip
we have detected at $\phi_{so}=0.6$--0.7. 
In an eccentric binary like \lsi, the strong tidal interaction between
the compact star and the Be disk would truncate the disk 
to a size smaller than the periastron distance if the viscosity 
parameter $\alpha\ll 1$ \citep{no01,oka+02}. As a result, the Be disk would 
be denser, which
has likely been observed in PSR B1259$-$63/LS~2883 when the pulsar passed
through periastron \citep{che+14}. However, \citet{zam+16} have recently
concluded that the compact star in \lsi\ passes through
the outer part of the disk at periastron (from their optical spectrosopic
study of the binary).  Therefore when the more extended or denser part of 
the Be disk rotates over the periastron region, 
it is not unreasonable to suggest that the disk possibly extends over 
the periastron substantially, causing the compact star or the emission
region to be totally inside the disk. The synchrotron emission
is thus significantly reduced (due to blocking/scattering), forming the dip.
However, no other available observational results appear to support
this dip forming possibility.  At X-rays, the very recent
analysis of all the archival X-ray imaging data has shown significant 
variations of $N_{\rm H}$ values, but the lack of observations 
during $\phi_{so}=0.6$--0.7 around periastron does not allow an 
assessment to be made.  In addition, we note that in
the case of PSR B1259$-$63/LS~2883, when the pulsar entered the Be disk region,
no significant $N_{\rm H}$ enhancement was observed \citep{che+14}.
Whether we would see significantly increased $N_{\rm H}$ is not clear.
In any case, X-ray observations of \lsi\ at particular phase ranges 
(i.e., $\phi_{so}=0.6$--0.7 and $\phi_{o}\simeq 0.27$) are of interest,
for the purpose of fully studying this binary.

The other possibility of forming a dip may be based on the scenario
of the compact star being a radio pulsar,
which has long been proposed for \lsi\ (e.g., \citealt{mt81,dub06}).
In this case, the dip would indicate the quench of the rotation-powered
activity or the state changes (see \citealt{tor+12}).
During $\phi_{so}\simeq 0.65$, larger amount of
matter from the Be disk may be captured by the pulsar and 
form an accretion disk.  There will be
 an relative angular momentum of the captured matter with respect to the 
pulsar. The circularization radius of the captured matter may be estimated 
as \citep{fkr02,tak+17}
\begin{eqnarray}
r_{circ}&\sim& \frac{r_{cap}^4\omega^2}{16GM_{N}}\sim 1.7\times 10^{10}\ {\rm cm}\nonumber \\
&\times& \left(\frac{M_{N}}{1.4\ M_{\odot}}\right)^3
\left(\frac{\omega}{5\cdot 10^{-7}\,{\rm rad~s^{-1}}}\right)^2
\left(\frac{v_{r}}{10^7\,{\rm cm~s^{-1}}}\right)^{-8}\nonumber ,
\end{eqnarray}
where $M_N$ is the neutron star mass, $\omega$ is the relative angular 
velocity, and $v_{r}$ is the relative velocity of the pulsar with respect 
to the disk matter. In addition, $r_{cap}\sim 2GM_{N}/v_r^2$ is 
the capture radius measured from the pulsar.  If we apply the standard 
Shakura-Sunyaev disk model, the disk matter will develop with a 
dynamical time scale of 
$\tau_d(r)\sim r/v_{d,r}\sim 15\alpha_{0.1}^{-4/5}\dot{M}_{17}^{-3/10}r_{10}^{3/4}{\rm days}$, where $\alpha_{0.1}$ is the 
viscous parameter in units of 0.1, suggesting 
the accreting matter will take several weeks to reach the pulsar 
after the capture event. It is also noted that 
the accreting matter will still exist even after the pulsar 
exits from the Be disk. Since the orbit period is close 
to the dynamical time scale of the accreting matter, 
the matter could remain entire orbit once the accretion disk 
forms around the pulsar. In Figure~\ref{fig:orbital_resolved_superorbital},
one notable feature is that the minumum flux of the superorbital modulation
during $\phi_{o}=0.5$--0.9 is also at $\phi_{so}=0.6-0.7$,
which could be explained by the switching-off of
the rotation-powered activity (or the state changes; \citealt{tor+12}) 
due to the formation of the accretion disk around the pulsar.

Once $\phi_{so}=0.65$ is tied to periastron, 
the fast/slow phase shifts around periastron/apastron are naturally explained. 
For example, examining the superorbital light curves in 
Figure~\ref{fig:orbital_resolved_superorbital}, we may conclude that
there are no significant phase shifts in orbital phases of 0.5--1.0. This
may be because of the relatively large distance between the disk and
the compact star around apastron; as a result, the modulation is not
sensitive to the precession of the disk. 
The overall superorbital modulation at different wavelengths can also 
be re-examined (see Figure~\ref{fig:orb}).
The so-called radio outburst actually starts
from $\phi_{so}=0.65$ and reaches the peak during 
$\phi_{so}=0.8$--1.0 \citep{gre02,tor+12}, and detailed modelling of 
8.3 GHz data suggests a launch of a density enhancement or shell in the Be disk
occurring around $\phi_{so}=0.6$ \citep{gn02}. 
The coincidence could be because after a stronger interaction 
between the compact star and the Be disk at $\phi_{so}=0.65$,
stronger outflows are driven from the system and
radio emission is largely enhanced.
Both the X-ray and GeV \gr\ peaks are at $\phi_{so}=0.1$--0.2
\citep{li+12,ack+13}, while the TeV \gr\ peak is rather uncertain,
with a large range of $\phi_{so}=0.1$--0.6 \citep{ahn+16}. 
According to the scenario we propose here, the X-ray and GeV \gr\ peaks
are when the non-axisymmetric Be disk points towards the apastron 
direction (cf., Figure~\ref{fig:orb}), which may
thus favor the pulsar scenario for \lsi.
In Figure~\ref{fig:orbital_resolved_superorbital}, we can see that 
the overall \gr\ peak during $\phi_{so}=0.1$--0.2 (see Figure 2 in 
\citealt{ack+13}) is because of 
the contribution from the emission peaks around apastron ($\phi_{o}=0.7$--0.9).
This feature implies that the ambient density around the compact star 
should not be a dominant factor deciding the emission intensity anymore; 
otherwise we would expect the emission peaks always around periastron. 
Some other factor, such as the rotational power of a working radio pulsar
(probably in form of a pulsar wind such as in the case of 
PSR B1259$-$63/LS~2883), would play a more important role in producing
X-ray and \gr\, emission. Due to the limited TeV data, it is not clear how
TeV emission is related to the scenario proposed here, but its
superorbital modulation has been suggested to be related to the changes between
the states of a propeller and a pulsar \citep{tor+12,ahn+16}.

\acknowledgements
This research made use of the High Performance Computing Resource in the Core
Facility for Advanced Research Computing at Shanghai Astronomical Observatory.
This research was supported by the National Program on Key Research 
and Development Project (Grant No. 2016YFA0400804) and
the National Natural Science Foundation
of China (11373055, 11403075, 11633007). Z.W. acknowledges the support by 
the CAS/SAFEA International Partnership Program for Creative Research Teams.
JT is supported by NSFC grants of Chinese Government
under 11573010 and U1631103.

\clearpage
\begin{table}
\tabletypesize{\footnotesize}
\tablewidth{240pt}
\setlength{\tabcolsep}{2pt}
\caption{Exponentially cutoff power-law fits for \lsi.}
\label{tab:likelihood}
\centering
\begin{tabular}{lccccc}
\hline
Data set & $>$0.1 GeV Flux & $\Gamma$ & E$_{c}$ & TS \\
 & (10$^{-7}$ photon cm$^{-2}$ s$^{-1}$) &  & (GeV) &  \\
\hline
Total data ($\Phi_{t}$) & 8.79$\pm$0.07 & 2.086$\pm$0.009 & 5.5$\pm$0.2 & 136681 \\
\hline
Periastron ($\Phi_{p}$) & 9.5$\pm$0.1 & 2.13$\pm$0.01 & 5.9$\pm$0.3 & 73272 \\
Dip ($\Phi_{d}$) & 8.8$\pm$0.3 & 2.27$\pm$0.02 & 16$\pm$4 & 3190 \\
Dip ($\Phi_{d}$, $<$5.5 GeV) & 8.7$\pm$0.3 & 2.15$\pm$0.03 & 4.2$\pm$0.5 & 3080 \\
\hline
Apastron ($\Phi_{a}$) & 8.13$\pm$0.09 & 2.04$\pm$0.01 & 5.1$\pm$0.3 & 63691 \\
\hline
\end{tabular}
\vskip 1mm
\end{table}

\clearpage
\begin{table}
\centering
\tabletypesize{\footnotesize}
\tablewidth{240pt}
\caption{\fermi\ LAT flux measurements of \lsi.}
\label{tab:spectra}
\tiny
\begin{tabular}{lcccc}
\hline
\diagbox{Band (GeV)}{Data set}& Total data  ($\Phi_{t}$) & Periastron  ($\Phi_{p}$) & Dip  ($\Phi_{d}$) & Apastron  ($\Phi_{a}$) \\
\hline
0.1--0.2 & 14.7$\pm$0.2 & 16.3$\pm$0.2 & 15$\pm$1 & 13.2$\pm$0.3 \\
0.2--0.5 & 14.2$\pm$0.1 & 14.9$\pm$0.2 & 13.2$\pm$0.8 & 13.6$\pm$0.2 \\
0.5--1.1 & 12.4$\pm$0.1 & 12.8$\pm$0.2 & 10.8$\pm$0.6 & 12.1$\pm$0.1 \\
1.1--2.5 & 9.1$\pm$0.1 & 9.1$\pm$0.2 & 7.8$\pm$0.6 & 9.1$\pm$0.2 \\
2.5--5.5 & 5.9$\pm$0.1 & 5.9$\pm$0.2 & 4.3$\pm$0.6 & 5.8$\pm$0.2 \\
5.5--12.2 & 2.8$\pm$0.1 & 3.0$\pm$0.2 & 3.0$\pm$0.8 & 2.6$\pm$0.2 \\
12.2--27.2 & 0.9$\pm$0.1 & 0.7$\pm$0.1 & 1.9$\pm$0.9 & 1.2$\pm$0.2 \\
27.2--60.5 & 0.5$\pm$0.1 & 0.5$\pm$0.2 & 2$\pm$1 & 0.4$\pm$0.1 \\
60.5--134.7 & 0.3$\pm$0.1 & 0.6 & 2.5 & 0.3$\pm$0.2 \\
134.7--300 & 0.5$\pm$0.2 & 0.6$\pm$0.4 & 5.8 & 0.4$\pm$0.3 \\
\hline
\end{tabular}
\vskip 1mm
\footnotesize{Fluxes are calculated from $E^2dN(E)/dE$ and in units 
of 10$^{-11}$ erg cm$^{-2}$ s$^{-1}$, while those values
without uncertainties are the 95$\%$ upper limits.}
\end{table}

\clearpage
\begin{table}
\tabletypesize{\footnotesize}
\tablewidth{240pt}
\setlength{\tabcolsep}{2pt}
\caption{Sinusoidal fits for the orbitally-resolved superorbital light curves of \lsi.}
\label{tab:fit_orbital_resolved_superorbital}
\centering
\begin{tabular}{cccccc}
\hline
Orbital phase & $A$/$10^{-7}$ & $C$/$10^{-7}$ & $\phi_{so, peak}$ & Significance \\
 & (photon cm$^{-2}$ s$^{-1}$) & (photon cm$^{-2}$ s$^{-1}$) & & of the dip ($\sigma$) \\
\hline
0.0--0.1 & 1.8$\pm$0.5 & 8.0$\pm$0.3 & 0.43$\pm$0.06 & 0.7 \\
0.1--0.2 & 2.1$\pm$0.8 & 8.5$\pm$0.5 & 0.47$\pm$0.07 & 3.2 \\
0.2--0.3 & 2.9$\pm$0.6 & 9.1$\pm$0.3 & 0.56$\pm$0.03 & 5.0 \\
0.3--0.4 & 1.3$\pm$0.5 & 9.2$\pm$0.3 & 0.71$\pm$0.05 & 4.0 \\
0.4--0.5 & 1.4$\pm$0.7 & 7.9$\pm$0.5 & 1.00$\pm$0.08 & 2.2 \\
0.5--0.6 & 4.1$\pm$0.9 & 6.5$\pm$0.6 & 0.14$\pm$0.04 & 0.9 \\
0.6--0.7 & 2.7$\pm$0.9 & 6.7$\pm$0.5 & 0.12$\pm$0.05 & 1.0 \\
0.7--0.8 & 3.2$\pm$0.8 & 6.5$\pm$0.5 & 0.11$\pm$0.04 & 1.3 \\
0.8--0.9 & 5.2$\pm$0.9 & 5.0$\pm$0.6 & 0.17$\pm$0.03 & 0.3 \\
0.9--0.0 & 3.3$\pm$0.5 & 6.3$\pm$0.3 & 0.26$\pm$0.02 & 1.2 \\
\hline
\end{tabular}
\vskip 1mm
\end{table}

\end{document}